\documentclass[10pt]{article}
\usepackage{fullpage,citesort,epsfig,graphics,amsbsy}
\usepackage{psfrag}
\usepackage[small]{caption}
\newcommand{\beq}{\begin{equation}}
\newcommand{\eeq}{\end{equation}}
\newcommand{\bea}{\begin{eqnarray}}
\newcommand{\eea}{\end{eqnarray}}

\newcommand{\Tr}{{\rm Tr}}
\newcommand{\be}{\begin{equation}}
\newcommand{\ee}{\end{equation}}
\newcommand{\bq}{\begin{eqnarray}}
\newcommand{\eq}{\end{eqnarray}}

\def\math{\mathsurround=0pt }
\def\leftrightarrowfill{$\math \mathord\gets \mkern-6mu \cleaders\hbox{$\mkern-2mu \mathord- \mkern-2mu$}\hfill
 \mkern-6mu \mathord\to$}
\def\overleftrightarrow#1{\vbox{\ialign{##\crcr
     \leftrightarrowfill\crcr\noalign{\kern-1pt\nointerlineskip}
     $\hfil\displaystyle{#1}\hfil$\crcr}}}

\newcommand{\bfs}{\boldsymbol}
\newcommand{\VEV}[1]{\langle#1\rangle}

\let\l=\lambda

 \def\bd{\begin{document}} \def\ed{\end{document}}
\def\ds{\documentstyle} \let\fr=\frac \let\bl=\bigl \let\br=\bigr
\let\Br=\Bigr \let\Bl=\Bigl
\let\bm=\bibitem
\let\na=\nabla
\let\pa=\partial \let\ov=\overline
\def\ft#1#2{{\textstyle{{\scriptstyle #1}\over {\scriptstyle #2}}}}
\def\fft#1#2{{#1 \over #2}}
\def\vp{\varphi}
\def\sst#1{{\scriptscriptstyle #1}}
\def\oneone{\rlap 1\mkern4mu{\rm l}}
\def\td{\tilde}
\def\wtd{\widetilde}
\def\dalemb#1#2{{\vbox{\hrule height .#2pt
        \hbox{\vrule width.#2pt height#1pt \kern#1pt
                \vrule width.#2pt}
        \hrule height.#2pt}}}
\def\square{\mathord{\dalemb{6.8}{7}\hbox{\hskip1pt}}}
\def\wtd{\widetilde}
\def\R{\rlap{\rm I}\mkern3mu{\rm R}}
\def\im{{\rm i}}
\def\tilg{\tilde{g}}
\def\tilF{\tilde{F}}
\def\tilA{\tilde{A}}
\def\varf{\varphi}
\def\tilf{\tilde{\phi}}
\def\tilh{\tilde{h}}
\def\rme{{\rm e}}
\def\ep{\epsilon}
\def\0{{(0)}}
\def\9{{(9)}}
\def\8{{(8)}}
\def\7{{(7)}}
\def\6{{(6)}}
\def\5{{(5)}}
\def\4{{(4)}}
\def\3{{(3)}}
\def\2{{(2)}}
\def\1{{(1)}}
\newcommand{\trace}{{\rm Tr}}
\newcommand{\ub}{\overline{U}}
\newcommand{\vb}{\overline{V}}
\newcommand{\uh}{\widehat{U}}
\newcommand{\vh}{\widehat{V}}
\newcommand{\ubh}{\overline{\widehat{U}}}
\newcommand{\vbh}{\overline{\widehat{V}}}
\newcommand{\lb}{\bar{\l}}
\newcommand{\Fb}{\overline{F}}
\newcommand{\Fh}{\widehat{F}}
\newcommand{\Fbh}{\overline{\widehat{F}}}
\newcommand{\Ab}{\overline{A}}
\newcommand{\Ah}{\widehat{A}}
\newcommand{\Abh}{\overline{\widehat{A}}}
\newcommand{\Gb}{\overline{G}}
\newcommand{\Gh}{\widehat{G}}
\newcommand{\Gbh}{\overline{\widehat{G}}}
\newcommand{\Pb}{\overline{P}}
\newcommand{\Ph}{\widehat{P}}
\newcommand{\Pbh}{\overline{\widehat{P}}}
\newcommand{\Qb}{\overline{Q}}
\newcommand{\Qh}{\widehat{Q}}
\newcommand{\Qbh}{\overline{\widehat{Q}}}
\newcommand{\Bb}{\overline{B}}
\newcommand{\Bh}{\widehat{B}}
\newcommand{\Bbh}{\overline{\widehat{B}}}
\newcommand{\fhns}{\hat{F}^{\rm (NS)}}
\newcommand{\fhrr}{\hat{F}^{\rm (RR)}}
\newcommand{\ahns}{\hat{A}^{\rm (NS)}}
\newcommand{\ahrr}{\hat{A}^{\rm (RR)}}
\newcommand{\hhrr}{\hat{H}^{\rm (RR)}}
\newcommand{\hchi}{\hat{\chi}}
\newcommand{\hphi}{\hat{\phi}}
\newcommand{\htau}{\hat{\tau}}
\newcommand{\cG}{{\cal G}}
\newcommand{\cGb}{\overline{{\cal G}}}
\newcommand{\cH}{{\cal H}}
\newcommand{\cP}{{\cal P}}
\newcommand{\cPb}{\overline{{\cal P}}}
\newcommand{\cQ}{{\cal Q}}
\newcommand{\cQb}{\overline{{\cal Q}}}
\newcommand{\cM}{{\cal M}}
\newcommand{\cN}{{\cal N}}
\newcommand{\cO}{{\cal O}}
\newcommand{\cD}{{\cal D}}
\newcommand{\cL}{{\cal L}}
\newcommand{\cA}{{\cal A}}
\newcommand{\cB}{{\cal B}}
\newcommand{\hg}{\hat{g}}
\newcommand{\cE}{{\cal E}}

\newcommand{\vpp}{\mbox{$\langle{\scriptstyle++}\rangle$}}
\newcommand{\vmp}{\mbox{$\langle{\scriptstyle-+}\rangle$}}
\newcommand{\vppp}{\mbox{$\langle{\scriptstyle+++}\rangle$}}
\newcommand{\vmpp}{\mbox{$\langle{\scriptstyle-++}\rangle$}}
\newcommand{\vpmp}{\mbox{$\langle{\scriptstyle+-+}\rangle$}}
\newcommand {\Kftw} {K^\land_{43}}
\newcommand {\Kttw} {K^\land_{32}}
\newcommand {\Ktow} {K^\land_{21}}
\newcommand {\Kofw} {K^\land_{14}}
\newcommand {\Kftv} {K^\lor_{43}}
\newcommand {\Kttv} {K^\lor_{32}}
\newcommand {\Ktov} {K^\lor_{21}}
\newcommand {\Kofv} {K^\lor_{14}}
\newcommand {\Po} {k^+_1}
\newcommand {\Pt} {k^+_0}
\newcommand {\Pth} {k^+_3}
\newcommand {\Pf} {k^+_2}
\newcommand {\Kp} {q^+}
\newcommand {\T} {T_1}
\newcommand {\Uu} {T_0}
\newcommand {\eS} {T_3}
\newcommand {\V} {T_2}
\newcommand {\invt} {(-t)}
\newcommand {\pref} {\frac{1}{8\pi^2}}
\newcommand {\rleft} {\Pt<\Kp<\Pth}
\newcommand {\rmid} {\Pth<\Kp<\Po}
\newcommand {\rright} {\Po<\Kp<\Pf}
\newcommand {\Anei} {A_{\land\land\lor\lor}}
\newcommand {\Aalt} {A_{\land\lor\land\lor}}
\newcommand {\Nc} {}
\newcommand{\eqe}{\!\!\! &=& \!\!\!}
\newcommand{\sme}{\!\!\! &\simeq& \!\!\!}
\newcommand{\sm}{\!\!\! &\sim& \!\!\!}
\newcommand{\sn}{\rm sn}
\newcommand{\dn}{\rm dn}
\newcommand{\cn}{\rm cn}
\usepackage{color}
\begin{document}

\setlength{\captionmargin}{20pt}
\begin{titlepage}
\begin{flushright}
\phantom{UFIFT-HEP-10-}
\end{flushright}

\vskip 2.5cm

\begin{center}
\begin{Large}
{\bf The Open String Regge Trajectory and\\ Its Field
Theory Limit\footnote{Supported 
in part by the Department
of Energy under Grant No. DE-FG02-97ER-41029.}}
\end{Large}

\vskip 2cm
{\large 
Francisco Rojas\footnote{E-mail  address: {\tt frojasf@phys.ufl.edu}} 
and Charles B. Thorn\footnote{E-mail  address: {\tt thorn@phys.ufl.edu}}
}
\vskip0.20cm
{\it Institute for Fundamental Theory\\
Department of Physics, University of Florida,
Gainesville FL 32611}
\vskip 1.0cm
\end{center}

\begin{abstract}\noindent
We study the properties of the leading Regge trajectory in open string
theory including the open string planar one-loop  corrections. With $SU(N)$
Chan-Paton factors, the sum over planar open string multi-loop diagrams
describes the 't Hooft limit $N\to\infty$ with $Ng_s^2$ fixed. 
Our motivation is to improve the understanding of open string theory
at finite $\alpha^\prime$ as
a model of gauge field theories. $SU(N)$ gauge theories in $D$
space-time dimensions are described by requiring
open strings to end on 
a stack of $N$ D$p$-branes of space-time dimension $D=p+1$.  
The large $N$ leading trajectory
$\alpha(t)=1+\alpha^\prime t+\Sigma(t)$ 
can be extracted, through order $g^2$,
from the  $s\to-\infty$ limit, at fixed $t$, 
of the four open string tree and planar 
loop diagrams. We analyze the $t\to0$ behavior with the result that
$\Sigma(t)\sim -Cg^2(-\alpha^\prime t)^{(D-4)/2}/(D-4)$.
This result precisely tracks the 1-loop Reggeized gluon of 
gauge theory in $D>4$ space-time dimensions. 
In particular, for $D\to4$ it reproduces the
known infrared divergences of gauge theory in 4 dimensions with a
Regge trajectory behaving as $-\ln(-\alpha^\prime t)$.
We also
study $\Sigma(t)$ in the limit $t\to-\infty$ and show that, when $D<8$, 
it behaves as
$\alpha^\prime t/(\ln(-\alpha^\prime t))^{\gamma}$, where $\gamma>0$ 
depends on $D$ and the number of massless scalars.
Thus, as long as $4<D<8$, the 1-loop correction stays small
relative to the tree trajectory for the whole range $-\infty<t<0$. 
Finally we present the results
of numerical calculations of $\Sigma(t)$ for all negative $t$.
\end{abstract}
\vfill
\end{titlepage}
\section{Introduction}
Ever since the early days of dual resonance models, 
it has been recognized that
quantum field theory can be recovered from those models as the zero slope
(infinite string tension) limit \cite{scherk,neveuscherk}. 
In more recent years this field/string relationship
has been exploited to motivate the AdS/CFT correspondence \cite{maldacenasole}, 
in which the closed strings remain stringy even in the zero-slope limit:
in this limit, closed superstring theory on an AdS$_5\times$S$_5$ 
background with a coincident stack of D3-branes 
is then conjectured to be equivalent to ${\cal N}=4$ supersymmetric
Yang-Mills theory. This conjecture is buttressed by the ability to do
explicit calculations at weak coupling via the gauge theory loop
expansion and at strong coupling via a semiclassical treatment of the
closed string on AdS.

However, gauge theories with less or no supersymmetry ($0\leq{\cal N}<4$) 
are also zero slope limits of corresponding open string theories. 
We think it is reasonable to
expect the techniques of string theory to deepen our 
dynamical understanding
of these theories too. Without maximal supersymmetry, quantum effects
typically break the conformal symmetry of the classical gauge theory,
linking the effective coupling to the scale of the process studied.
The coupling constant ceases to be an adjustable parameter:
at high momenta the coupling is weak and field theoretic
perturbation theory is applicable. However, at low momenta the
coupling is at least of order 1, though not necessarily
very large. Unlike the conformal theories,
this means that semi-classical string methods, controlled by
the infinite coupling limit, cannot
be reliably applied to the putative closed string theory which
might represent such a gauge theory. 

Under these circumstances, we think a useful line of attack is to replace
the field theory with its corresponding open string theory,
delaying the $\alpha^\prime\to0$ limit to the end of the calculation.
In other words, we seek to resolve nonperturbative issues in
gauge theory by summing open string multiloop diagrams instead
of field theoretic multi-loop diagrams. Since closed string
intermediate states inevitably participate in nonplanar
open string diagrams, we expect this approach to be most
fruitful for the summation of planar open string diagrams only.
That is, we expect this approach to teach us the most about gauge theories
in 't Hooft's large $N$ limit \cite{thooftlargen}.

Keeping $\alpha^\prime>0$ can be viewed as a regularization procedure
for the program developed in \cite{bardakcit,thornsheet} to represent
and sum the planar diagrams of gauge theory on a lightcone worldsheet lattice
\cite{gilest}.
In this representation the gluon is replaced by a kind of topological
open string whose only physical energy eigenstate 
is the spin one massless gluon. In practice, however, the lightcone
lattice produces ultraviolet artifacts, the cancellation of which
requires worldsheet counterterms beyond coupling and wave function
renormalization \cite{chakrabartiqt,chakrabartiqt2}. It
seems likely that more and more 
such counterterms will be needed at each order of perturbation theory.
Since the excitations of the topological string cancel
in the worldsheet path integral, they have no mitigating effect on the
field theoretic ultraviolet divergences. In contrast, with $\alpha^\prime>0$
the open string excitations are real, and they 
do soften the ultraviolet divergences.

With this application in mind, it behooves us to
understand perturbative open string theory as well
as we understand perturbative gauge theory. In this article we begin this
task with a study of one loop open string physics. As a 
concrete example we focus on the one loop correction to the
leading open string Regge trajectory\footnote{This choice is
inspired by recent work on the gluon Regge trajectory 
in the context of the AdS/CFT correspondence \cite{schnitzer}.}. 
Writing $\alpha(t)
=\alpha^\prime t + 1 + \Sigma(t)$ with $\Sigma(t)=O(g^2)$, 
we see that the zero slope
limit of $\Sigma$ should just describe the known reggeization of
the gauge particle (which we call the gluon in this article).
With our new viewpoint, we prefer to think of $\alpha(t)$ as
a fundamental physical quantity in open string theory with
$\alpha^\prime>0$ and fixed. Then the reggeized gauge theory
gluon is simply the part of this trajectory near $t=0$. 
But the trajectory away from $t=0$ is also significant to open string physics.
Therefore we study both the small and large $t$ behavior of
$\Sigma$ analytically. And we also carry out numerical calculations
for the whole range of negative $t$.

We organize the paper as follows. In section 2 we recall the
one loop amplitude for the Neveu-Schwarz model \cite{neveuschwarz,neveust}, 
as calculated long ago \cite{goddardw}. In order to obtain
the gauge theories as zero slope limits, we adapt that calculation in
several ways. First of all we work in the critical dimension
throughout: the interior points of the
open string are allowed to move in 10 spacetime
dimensions. This greatly simplifies the elimination of 
unphysical states from the loop \cite{browert,goddardt}.
Open string tachyons are eliminated by working in the
even G-parity subspace of open string states \cite{mandelstam}
and projecting out odd G-parity states in the loop. We also
require the open string to end on D$p$-branes, with $p=D-1$
and $D$ the spacetime dimension of the branes. Finally, when $D<10$ we
arrange for only $S\leq10-D$ massless scalars to circulate in the
loop.\footnote{For simplicity, in the following we
shall use the nonabelian D-brane projection proposed in \cite{thornnonabelian}.
Another approach is to retain only states even under $b_r^A, a_n^A\to
-b_r^A,-a_n^A$ where $D+S<A\leq10$. In this case, the following changes
occur in $Z$:
\bea
(1-w^{1/2})^{10-D-S}{\prod(1+w^r)^8\over\prod(1-w^n)^8}\to
{1\over2}\left[{\prod(1+w^r)^8\over\prod(1-w^n)^8}
+{\prod(1+w^r)^{D-2+S}\prod(1-w^r)^{10-D-S}\over\prod(1-w^n)^{D-2+S}
\prod(1+w^n)^{10-D-S}}\right]\nonumber\\
(1+w^{1/2})^{10-D-S}{\prod(1-w^r)^8\over\prod(1-w^n)^8}\to
{1\over2}\left[{\prod(1-w^r)^8\over\prod(1-w^n)^8}
+{\prod(1-w^r)^{D-2+S}\prod(1+w^r)^{10-D-S}\over\prod(1-w^n)^{D-2+S}
\prod(1+w^n)^{10-D-S}}\right]\nonumber
\eea }
This extends the applicability of our methods to pure gauge theory, $S=0$.
These various projections
are summarized by writing out the partition functions in
the ${\cal F}_2$ open string state space \cite{neveust}
\bea
Z&=&\Tr w^{L_0-1/2}={w^{-1/2}\over2}\left[{(1-w^{1/2})^{10-D-S}\over
\ln^{D/2}w}{\prod(1+w^r)^8\over\prod(1-w^n)^8}-
{(1+w^{1/2})^{10-D-S}\over
\ln^{D/2}w}{\prod(1-w^r)^8\over\prod(1-w^n)^8}\right]
\eea
$n=1,2,\cdots$, $r=1/2,3/2,\cdots$
Working with these
results, and following the methods of \cite{kikkawasv,alessandrini,ottope}, 
we extract the one-loop correction $\Sigma$ to the open string
Regge trajectory in Section 3. 

We then turn to a detailed study of the properties of $\Sigma$.
Section 4 is devoted to a study of the small $t$ behavior of $\Sigma$.
This limit exposes
the infrared structure of the open string theory, which is 
identical to that of the corresponding gauge theory. As long
as $D>4$ infrared divergences are absent, and we confirm that
$\Sigma(0)=0$, a reflection of the zero mass of the gluon.  
In Section 5 we study the large $t$ behavior of $\Sigma(t)$, which exposes
the ultraviolet structure of open string theory. Although this
is not the same as the ultraviolet structure of the gauge theory,
it shares a common property as long as $D<8$: ultraviolet divergences
can be absorbed in coupling renormalization. For $D\geq8$ 
subleading divergences require renormalization of the Regge
slope parameter $\alpha^\prime$. 
In Section 6 we
describe our numerical work, illustrating it with graphs
of $\Sigma(t)$ in various regimes. We close with 
discussion and concluding remarks in Section 7. 
\section{The One Loop Correction}
The open string coupling $g$ will be normalized in this
paper so that in the zero-slope limit it is 
related to the QCD strong coupling
$g_s$ by $\alpha_s N=g_s^2N/4\pi=g^2/2\pi$. Thus $g$ will be fixed in the
large $N$ limit.
Then the properly normalized  
$M$ gluon open string one loop planar amplitude
for the even G-parity Neveu-Schwarz (NS+) model \cite{goddardw,metsaevt} is
$(g\sqrt{2\alpha^\prime})^M$ times 
\bea
{\cal M}_M={1\over2}({\cal M}_M^+-{\cal M}_M^-)\label{emoneloop}
\eea
where
\bea
{\cal M}_M^\pm
&=&\int {dw\over w}  
\prod_{i=2}^M {dy_i\over y_i}w^{-1/2}
\left({-1\over4\pi\alpha^\prime\ln w}\right)^{D/2}
\exp\left\{\alpha^\prime
\sum_{i< j}k_i\cdot k_j{\ln^2{y_i/y_j}\over\ln w}\right\}
\nonumber\\
&&(1\mp w^{1/2})^{10-D-S}
\VEV{{\hat{\cal P}}(y_1)\cdots{\hat{\cal P}}(y_M)}^\pm{\prod_r(1\pm w^r)^8
\over\prod_n(1-w^n)^8}
\prod_{i<j}\left[2i{\theta_1\left(-i\ln\sqrt{y_i/ y_j},\sqrt{w}\right)
\over
\theta_1^\prime(0,\sqrt{w})}
\right]^{2\alpha^\prime k_i\cdot k_j}
\label{empee}
\eea
where we use the notation and conventions
of \cite{thornsubcritical}. The integration range for the 
Koba-Nielsen variables $y_i$ is given by
\bea
0<w<y_M<y_{M-1}<\cdots <y_2<y_1=1\;.
\eea
We have adapted the standard planar
open string one loop calculation in 10 space-time
dimensions to open strings ending on a stack of
$N$ coincident $Dp$-branes for $p=D-1$.
In the planar one-loop calculation, 
this simply amounts to integrating over only
the first $D$ components of the loop momentum and setting the
remaining components to zero.
We also arrange that there are $S\leq 10-D$
massless scalars, achieved here by the extra factors of
$(1\mp w^{1/2})$ as dictated by the non-abelian D-brane
projection described in \cite{thornnonabelian}. 
The factors involving the Jacobi $\theta_1$
function have the infinite product representation
\bea
\prod_{i<j}y_j^{2\alpha^\prime k_i\cdot k_j}
\prod_{i<j}\left[2i{\theta_1\left(-i\ln\sqrt{y_i/y_j},\sqrt{w}\right)
\over
\theta_1^\prime(0,\sqrt{w})}
\right]^{2\alpha^\prime k_i\cdot k_j}&=&\nonumber\\
&&\hskip-1.5in\prod_{i<j}\left[\left(1-{y_j\over y_i}\right)\prod_n
{\left(1-w^n{y_i/y_j}\right)\left(1-w^n{y_j/y_i}\right)
\over(1-w^n)^2}\right]^{2\alpha^\prime k_i\cdot k_j}\;.
\eea
The gluon vertex operator is $V=e^{i k\cdot x}(\epsilon\cdot{\cal P}
+\sqrt{2\alpha^\prime}k\cdot H\epsilon\cdot H)\equiv e^{i k\cdot x}
{\hat{\cal P}}$.
The $\VEV{\cdots}$ is a correlator of a finite number of
${\cal P}$ and $H$ worldsheet fields determined by its Wick expansion
with the following contraction rules
\bea
\VEV{{\cal P}(y_l)}&=&\sqrt{2\alpha^\prime}\sum_i k_i\left[
-{\ln(y_i/y_l)\over\ln w}+{1\over2}{y_i+y_l\over y_l-y_i}
+\sum_{n=1}^\infty\left({y_iw^n\over y_l-y_iw^n}-{y_lw^n\over y_i-y_lw^n}
\right)\right]\nonumber\\
\VEV{{\cal P}^\mu(y_i){\cal P}^\nu(y_l)}&=&\VEV{{\cal P}^\mu(y_i)}
\VEV{{\cal P}^\nu(y_l)}
+\eta^{\mu\nu}\left[
-{1\over \ln w}+{y_i y_l\over(y_i-y_l)^2}
+\sum_{n=1}^\infty\left({y_i y_lw^n\over(y_l-y_iw^n)^2}
+{y_i y_lw^n\over(y_i-y_lw^n)^2}\right)\right]\nonumber\\
\VEV{H^\mu(y_i)H^\nu(y_j)}^+&=&\eta^{\mu\nu}
\sum_r{(y_j/y_i)^r+(wy_i/y_j)^r\over
1+w^r}\nonumber\\
\VEV{H^\mu(y_i)H^\nu(y_j)}^-&=&\eta^{\mu\nu}
\sum_r{(y_j/y_i)^r-(wy_i/y_j)^r\over
1-w^r} .
\eea
The $\pm$ superscript on the $H$ contractions distinguishes the
two types of traces over the $b_r$ oscillators: for $+$ odd
and even G-parity states contribute with the same sign, whereas
for $-$ they contribute with opposite signs. In the ${\cal F}_2$
picture, the difference of the two traces projects out the odd 
G-parity states.

Finally we present the one-loop amplitude in 
the natural cylinder variables, $\theta_i=\pi\ln y_i/\ln w$
and $\ln q=2\pi^2/\ln w$,
\bea
{\cal M}_M^+&=&
2^M\left({1\over8\pi^2\alpha^\prime}\right)^{D/2}
\int \prod_{k=2}^M {d\theta_k}\int_0^1{dq\over q}
\left({-\pi\over\ln q}\right)^{(10-D)/2}
q^{-1}(1-w^{1/2})^{10-D-S}\nonumber\\
&&\qquad
{\prod_r(1+q^{2r})^{8}\over\prod_n(1-q^{2n})^{8}}
\prod_{l<m}\left[\psi(\theta_m-\theta_l,q)
\right]^{2\alpha^\prime k_l\cdot k_m}
\VEV{{\hat{\cal P}}_1{\hat{\cal P}}_2\cdots{\hat{\cal P}}_M}^+\nonumber\\
\nonumber\\
{\cal M}_M^-&=&
2^M\left({1\over8\pi^2\alpha^\prime}\right)^{D/2}
\int \prod_{k=2}^M {d\theta_k}\int_0^1{dq\over q}
\left({-\pi\over\ln q}\right)^{(10-D)/2}
2^{4}(1+w^{1/2})^{10-D-S}\nonumber\\
&&\qquad
{\prod_n(1+q^{2n})^{8}\over\prod_n(1-q^{2n})^{8}}
\prod_{l<m}\left[\psi(\theta_m-\theta_l,q)
\right]^{2\alpha^\prime k_l\cdot k_m}
\VEV{{\hat{\cal P}}_1{\hat{\cal P}}_2\cdots{\hat{\cal P}}_M}^-\label{empee-}\\
\psi(\theta,q)&=&\sin{\theta}\prod_n{(1-q^{2n}e^{2i\theta})
(1-q^{2n}e^{-2i\theta})\over(1-q^{2n})^2}\nonumber\\
{\hat{\cal P}}&=& \epsilon\cdot
{\cal P}+\sqrt{2\alpha^\prime}k\cdot H\epsilon\cdot H ,
\nonumber\eea
where the average $\VEV{\cdots}$ is evaluated with contractions:
\bea
\VEV{{\cal P}_l}&=&\sqrt{2\alpha^\prime}\sum_i k_i
\left[{1\over2}\cot{\theta_{il}}
+\sum_{n=1}^\infty{2q^{2n}\over1-q^{2n}}
\sin 2n\theta_{il}\right]\\
\VEV{{\cal P}_i{\cal P}_l}
-\VEV{{\cal P}_i}\VEV{{\cal P}_l}
&=&{1\over4}\csc^2{\theta_{il}}
-\sum_{n=1}^\infty n{2q^{2n}\over1-q^{2n}}
\cos 2n\theta_{il}\\
\VEV{H_iH_j}^+&\equiv&\chi_+(\theta_{ji})
={1\over2\sin\theta_{ji}}
-2\sum_r{q^{2r}\sin 2r\theta_{ji}\over
1+q^{2r}}={1\over2}\theta_2(0)\theta_4(0){\theta_3(\theta_{ji})\over
\theta_1(\theta_{ji})}\\
\VEV{H_iH_j}^-&\equiv&\chi_-(\theta_{ji})
={\cos\theta_{ji}\over2\sin\theta_{ji}}
-2\sum_n{q^{2n}\sin 2n\theta_{ji}\over1+q^{2n}} 
={1\over2}\theta_3(0)\theta_4(0){\theta_2(\theta_{ji})\over
\theta_1(\theta_{ji})}\;.
\eea
We have abbreviated $\theta_{ji}=\theta_j-\theta_i$
and we have again suppressed space-time indices.
Finally  the range of integration is
\bea
0=\theta_1<\theta_2<\cdots<\theta_N<\pi .
\eea
In these formulas $r$ ranges over positive half odd integers,
$n$ over positive integers, and $l,m \in [1,\cdots, M]$.  

The worst divergence in the $q$ integration near $q=0$, the $q^{-2}$
behavior in ${\cal M}^+$, can be cancelled by the Neveu-Scherk
counterterm \cite{neveuscherkrenorm}
\bea
{\cal C}_M&=&
2^M\left({1\over8\pi^2\alpha^\prime}\right)^{D/2}
\int \prod_{k=2}^M {d\theta_k}\int_0^1{dq\over q^2}
\left({-\pi\over\ln q}\right)^{(10-D)/2}
(1-w^{1/2})^{10-D-S}\nonumber\\
&&\qquad
{\prod_r(1+q^{2r})^{8}\over\prod_n(1-q^{2n})^{8}}
\prod_{l<m}\left[\sin{\theta_m-\theta_l}
\right]^{2\alpha^\prime k_l\cdot k_m}
\VEV{{\hat{\cal C}}_1{\hat{\cal C}}_2\cdots{\hat{\cal C}}_M}\label{nscounter}
\eea
where the average is evaluated with the contractions
\bea
\VEV{{\cal P}_l}_C&=&\sqrt{2\alpha^\prime}\sum_i k_i
\left[{1\over2}\cot{\theta_{il}}\right],\qquad 
\VEV{{\cal P}_i{\cal P}_l}_C
-\VEV{{\cal P}_i}_C\VEV{{\cal P}_l}_C
={1\over4}\csc^2{\theta_{il}}\nonumber\\
\VEV{H_iH_j}_C
&=&{1\over2\sin\theta_{ji}}\nonumber
\eea
As shown long ago \cite{neveuscherkrenorm},
when the $\theta$ integrals are regulated by temporarily suspending
momentum conservation $\sum_i k_i=P$ and analytically continuing to
$P=0$ \cite{goddardreg,neveuscherkrenorm}, the Neveu-Scherk counterterm 
simply goes to the tree amplitude. This establishes that this
leading divergence can be absorbed in renormalization of the
coupling constant. If the brane space-time dimension $D<8$,
there are no subleading ultraviolet divergences. Furthermore if $D>4$
there are no infrared divergences: after coupling 
renormalization the loop integral is completely
finite for $D=5,6,7$, and has only infrared divergences for $D\leq4$. 

As we have explained the $\theta$ integrals are well-defined
through analytic continuation from $P\neq0$. However, it is
useful in the following to be able to deal 
with well-defined expressions at $P=0$.
This can be accomplished in an unambiguous way by taking
$P\neq0$ and then subtracting and adding the
Neveu-Scherk counterterm: $I(P)=(I(P)-C(P))+C(P)$. Then it is safe to 
take $P\to0$ in the  first two terms. The $P\to0$ limit of
the last term $C(P)$ can be carefully studied, to trace the effects
of the divergence. 
\section{The Open String Regge Trajectory}
In general, the correction to the tree level Regge trajectory can be read off
from the large $s$ at fixed $t$ behavior of the one loop
amplitude \cite{kikkawasv,alessandrini,ottope}. 
Assuming Regge behavior of the exact amplitude,
\bea
(\beta(t)+\delta\beta)s^{\alpha(t)+\delta\alpha}
\approx\beta s^\alpha +\delta\alpha \beta s^\alpha \ln s 
+\delta\beta s^\alpha ,
\eea
$\delta\alpha$ is just the coefficient of $\beta s^\alpha \ln s$
in the one loop amplitude. 

In our case, the large $s$ behavior of ${\cal M}_M^\pm$ is controlled by
the region $\theta_{23},\theta_{41}\approx0$ or $\theta_2\approx\theta_3$
and $\theta_4\approx\pi$. The polarization factors of the Regge contribution
to the tree amplitude are 
$\epsilon_2\cdot\epsilon_3 \epsilon_1\cdot\epsilon_4$, 
so we pick out those terms in the correlator
\bea
\VEV{{\hat{\cal P}}_1{\hat{\cal P}}_2{\hat{\cal P}}_3{\hat{\cal P}}_4}
&\to&\epsilon_2\cdot\epsilon_3 \epsilon_1\cdot\epsilon_4\bigg(
\VEV{{{\cal P}}_2{{\cal P}}_3}\VEV{{{\cal P}}_1{{\cal P}}_4}-\VEV{{{\cal P}}_2{{\cal P}}_3}\VEV{H_1H_4}^22\alpha^\prime k_1\cdot k_4\nonumber\\
&&\hskip-1in
-\VEV{{{\cal P}}_1{{\cal P}}_4}\VEV{H_2H_3}^22\alpha^\prime k_2\cdot k_3
+4\alpha^{\prime2}\VEV{H_2H_3}\VEV{H_1H_4}
\VEV{k_1\cdot H_1 k_2\cdot H_2 k_3\cdot H_3 k_4\cdot H_4}\bigg)
\nonumber\\
&\to&\epsilon_2\cdot\epsilon_3 \epsilon_1\cdot\epsilon_4\bigg(
\VEV{{{\cal P}}_2{{\cal P}}_3}\VEV{{{\cal P}}_1{{\cal P}}_4}-\VEV{{{\cal P}}_2{{\cal P}}_3}\VEV{H_1H_4}^22\alpha^\prime k_1\cdot k_4\nonumber\\
&&\hskip-1in
-\VEV{{{\cal P}}_1{{\cal P}}_4}\VEV{H_2H_3}^22\alpha^\prime k_2\cdot k_3
+4\alpha^{\prime2}\VEV{H_2H_3}\VEV{H_1H_4}
(k_1\cdot k_2k_3\cdot k_4\VEV{H_1 H_2}\VEV{H_3H_4}\nonumber\\
&&-k_1\cdot k_3k_2\cdot k_4\VEV{H_1 H_3}\VEV{H_2H_4}
+k_1\cdot k_4k_2\cdot k_3\VEV{H_2 H_3}\VEV{H_1H_4}\bigg)
\nonumber\\
&\to&\epsilon_2\cdot\epsilon_3 \epsilon_1\cdot\epsilon_4\bigg(
(\VEV{{{\cal P}}_2{{\cal P}}_3}+\alpha^\prime t\VEV{H_2H_3}^2)(
\VEV{{{\cal P}}_1{{\cal P}}_4}+\alpha^\prime t\VEV{H_1H_4}^2)\nonumber\\
&&\hskip-1in
\VEV{H_2H_3}\VEV{H_1H_4}
(\alpha^{\prime2}s^2\VEV{H_1 H_2}\VEV{H_3H_4}
-\alpha^{\prime2}(s+t)^2\VEV{H_1 H_3}\VEV{H_2H_4}\bigg)
\eea
In the limit $\theta_{32}\to0,\theta_{41}\to\pi$ we have:
\bea
\VEV{{{\cal P}}_1{{\cal P}}_4}&\sim&{1\over4(\pi-\theta_4)^2},\qquad
\VEV{{{\cal P}}_2{{\cal P}}_3}\sim{1\over4(\theta_3-\theta_2)^2}\nonumber\\
\VEV{H_1 H_4}^\pm&\sim&{\pm 1\over2(\pi-\theta_4)},\qquad 
\VEV{H_2 H_3}^\pm\sim{1\over2(\theta_3-\theta_2)},\qquad\VEV{H_1 H_3}^\pm
=\chi_\pm(\theta_3)\nonumber\\
\VEV{H_1 H_2}^\pm&=&\chi_\pm(\theta_3-\theta_{32})
\sim\chi_\pm(\theta_3)-\theta_{32}\chi_\pm^{\prime}(\theta_3)
+{\theta_{32}^2\over2}\chi_\pm^{\prime\prime}(\theta_3)\nonumber\\
\VEV{H_3 H_4}^\pm&=&
\chi_\pm(\pi-\theta_3-(\pi-\theta_4))=\pm
\chi_\pm(\theta_3+(\pi-\theta_4))\nonumber\\
&\sim&\pm\left(
\chi_\pm(\theta_3)+(\pi-\theta_4)\chi_\pm^{\prime}(\theta_3)
+{(\pi-\theta_4)^2\over2}\chi_\pm^{\prime\prime}(\theta_3)\right)\nonumber\\
\VEV{H_2 H_4}^\pm&=&
\chi_\pm(\pi-\theta_3+\theta_{32}-(\pi-\theta_4))=\pm
\chi_\pm(\theta_3+(\pi-\theta_4)-\theta_{32})\nonumber\\
&\sim&\pm\left(
\chi_\pm(\theta_3)+(\pi-\theta_4-\theta_{32})\chi^{\pm\prime}(\theta_3)
+{(\pi-\theta_4-\theta_{32})^2\over2}\chi_\pm^{\prime\prime}(\theta_3)\right)
\nonumber\\
&&\hskip-1in \VEV{H_1 H_2}\VEV{H_3H_4}-\VEV{H_1 H_3}\VEV{H_2H_4}\ \sim\ 
\pm\theta_{32}(\pi-\theta_4)(\chi_\pm(\theta_3)
\chi_\pm^{\prime\prime}(\theta_3)
-\chi_\pm^{\prime2}(\theta_3))
\eea
Putting these forms into the correlator we have 
\bea
\VEV{{\hat{\cal P}}_1{\hat{\cal P}}_2{\hat{\cal P}}_3{\hat{\cal P}}_4}
&\sim&\epsilon_2\cdot\epsilon_3 \epsilon_1\cdot\epsilon_4
\bigg({(1+\alpha^\prime t)^2\over
16\theta_{32}^2(\pi-\theta_4)^2}+{1\over4}
(\alpha^\prime s)^2(\chi_\pm(\theta_3)
\chi_\pm^{\prime\prime}(\theta_3)
-\chi_\pm^{\prime2}(\theta_3))\nonumber\\
&&\hskip2in -\alpha^{\prime2}\chi_\pm^{2}(\theta_3)
{2st+t^2\over4\theta_{32}(\pi-\theta_4)}\bigg)
\eea
The $s$ dependent factors in the four string 1-loop diagram involve
the combination:
\bea
{\psi(\theta_{43})\psi(\theta_{21})\over
\psi(\theta_{42})\psi(\theta_{31})}
&=&
{\psi(\theta_3+(\pi-\theta_{4}))
\psi(\theta_3-\theta_{32})\over
\psi(\theta_3-\theta_{32}+(\pi-\theta_4))\psi(\theta_{31})}\nonumber\\
&\sim&
{(\psi+(\pi-\theta_{4})\psi^\prime+(\pi-\theta_4)^2
\psi^{\prime\prime}/2)(\psi-\theta_{32}\psi^\prime+\theta_{32}^2
\psi^{\prime\prime}/2)\over
(\psi+(\pi-\theta_{4}-\theta_{32})\psi^\prime+(\pi-\theta_4-\theta_{32})^2
\psi^{\prime\prime}/2)\psi}
\nonumber\\
&\sim&
\exp\left\{-\theta_{32}(\pi-\theta_4)\left({\psi^{\prime2}\over\psi^2}
-{\psi^{\prime\prime}\over\psi}\right)\right\}=
\exp\{\theta_{32}(\pi-\theta_4)(\ln\psi)^{\prime\prime}\}\nonumber\\
\left({\psi(\theta_{43})\psi(\theta_{21})\over
\psi(\theta_{42})\psi(\theta_{31})}\right)^{-\alpha^\prime s}
&\sim&\exp\{-\alpha^\prime s\ \theta_{32}(\pi-\theta_4)
(\ln\psi)^{\prime\prime}\}
\eea
Meanwhile the $t$ dependence is given by the factor
\bea
\left({\psi(\theta_{41})\psi(\theta_{32})\over
\psi(\theta_{42})\psi(\theta_{31})}\right)^{-\alpha^\prime t}
&\sim&\left({\theta_{32}
(\pi-\theta_4)\over \psi^2(\theta_3)}\right)^{-\alpha^\prime t}
\eea
Taking $s\to-\infty$, the $\theta_{32},\theta_4\equiv\pi-{\hat\theta}_4
$ integrals are dominated by
small $\theta_{23}\approx0$, ${\hat\theta}_4\approx0$, and hence can be
done by using
\bea
\int_0^\epsilon d\xi\int_0^\epsilon d\eta (\xi\eta)^a e^{-\xi\eta \kappa}
&=&\int_0^\epsilon {d\xi\over\xi}\int_0^{\xi\epsilon} d\eta \eta^a
e^{-\eta\kappa}=\kappa^{-a-1}\int_0^{\epsilon^2\kappa}
d\eta \eta^a \ln{\epsilon^2\kappa\over\eta}
e^{-\eta\kappa}\nonumber\\
&\sim& \Gamma(a+1) \kappa^{-a-1} \ln\kappa,\qquad \kappa\to\infty
\eea
We use this formula for $\kappa=(-\alpha^\prime s)(-[\ln\psi]^{\prime\prime})$
and $a=-\alpha^\prime t -2,-\alpha^\prime t -1,-\alpha^\prime t$:
\bea
{\cal M}_4^+&\sim&
16\epsilon_2\cdot\epsilon_3 \epsilon_1\cdot\epsilon_4
\left({1\over8\pi^2\alpha^\prime}\right)^{D/2}
\int_0^1{dq\over q^2}\left({-\pi\over\ln q}\right)^{(10-D)/2}
(1-w^{1/2})^{10-D-S}{\prod_r(1+q^{2r})^{8}\over\prod_n(1-q^{2n})^{8}}
\nonumber\\
&&\int d\theta_3\int_0^\epsilon d\theta_{32}d{\hat\theta}_4
\left({\theta_{32}
{\hat\theta}_4\over \psi^2(\theta_3)}\right)^{-\alpha^\prime t}
\exp\{-\alpha^\prime s\ \theta_{32}{\hat\theta}_4
(\ln\psi)^{\prime\prime}\}
\nonumber\\
&&
\bigg({(1+\alpha^\prime t)^2\over
16\theta_{32}^2{\hat\theta}_4^2}+{1\over4}(\alpha^\prime s)^2(\chi_+(\theta_3)
\chi_+^{\prime\prime}(\theta_3)
-\chi_+^{\prime2}(\theta_3))-\alpha^{\prime2}\chi_+^{2}(\theta_3)
{2st+t^2\over4\theta_{32}{\hat\theta}_4}\bigg)\nonumber\\
&\sim&
4\epsilon_2\cdot\epsilon_3 \epsilon_1\cdot\epsilon_4
\left({1\over8\pi^2\alpha^\prime}\right)^{D/2}
\int_0^1{dq\over q^2}\left({-\pi\over\ln q}\right)^{(10-D)/2}
(1-w^{1/2})^{10-D-S}{\prod_r(1+q^{2r})^{8}\over\prod_n(1-q^{2n})^{8}}
\nonumber\\
&&\Gamma(-\alpha^\prime t)(-\alpha^\prime s)^{1+\alpha^\prime t}
\ln(-\alpha^\prime s)\ \int d\theta_3(-\psi^2(\theta_3)
[\ln\psi]^{\prime\prime})^{\alpha^\prime t}
\nonumber\\
&&
\bigg(-{1\over4}{(1+\alpha^\prime t)}[-\ln\psi]^{\prime\prime}-
\alpha^\prime t
{\chi_+(\theta_3)\chi_+^{\prime\prime}(\theta_3)
-\chi_+^{\prime2}(\theta_3)\over[-\ln\psi]^{\prime\prime}}
+{2\alpha^\prime t}
\chi_+^{2}(\theta_3)\bigg)\nonumber\\
{\cal M}_4^-
&\sim&
4\epsilon_2\cdot\epsilon_3 \epsilon_1\cdot\epsilon_4
\left({1\over8\pi^2\alpha^\prime}\right)^{D/2}
\int_0^1{dq\over q}2^{4}\left({-\pi\over\ln q}\right)^{(10-D)/2}
(1+w^{1/2})^{10-D-S}{\prod_n(1+q^{2n})^{8}\over\prod_n(1-q^{2n})^{8}}
\nonumber\\
&&\Gamma(-\alpha^\prime t)(-\alpha^\prime s)^{1+\alpha^\prime t}
\ln(-\alpha^\prime s)\ \int d\theta_3(-\psi^2(\theta_3)
[\ln\psi]^{\prime\prime})^{\alpha^\prime t}
\nonumber\\
&&
\bigg({-{1\over4}(1+\alpha^\prime t)}[-\ln\psi]^{\prime\prime}
-\alpha^\prime t
{\chi_-(\theta_3)\chi_-^{\prime\prime}(\theta_3)
-\chi_-^{\prime2}(\theta_3)\over[-\ln\psi]^{\prime\prime}}
+{2\alpha^\prime t}
\chi_-^{2}(\theta_3)\bigg)
\eea
The total planar 1-loop amplitude is given by
\bea
{\cal M}^{\rm1~loop}_4&=&(g\sqrt{2\alpha^\prime})^4
{{\cal M}^{+}_4-{\cal M}^{-}_4\over 2}.
\eea
In order to extract the correction to the Regge trajectory we need to
know the Regge behavior of the tree:
\bea
{\cal M}^{\rm Tree}&\sim&-2g^2
\epsilon_2\cdot\epsilon_3 \epsilon_1\cdot\epsilon_4 
\Gamma(-\alpha^\prime t)(-\alpha^\prime s)^{1+\alpha^\prime t}
\eea
Removing the tree factors and the $\ln(-\alpha^\prime s)$ from the
asymptotic loop amplitude, we define
\bea
\Sigma^{+}&=&-{8g^2\alpha^{\prime2-D/2}
\over(8\pi^2)^{D/2}}
\int_0^1{dq\over q^2}\left({-\pi\over\ln q}\right)^{(10-D)/2}
(1-w^{1/2})^{10-D-S}{\prod_r(1+q^{2r})^{8}\over\prod_n(1-q^{2n})^{8}}
\nonumber\\
&&\hskip-.28in  \int_0^{\pi} d\theta\bigg((-\psi^2(\theta)
[\ln\psi]^{\prime\prime})^{\alpha^\prime t}
{\alpha^\prime t\over[\ln\psi]^{\prime\prime}}
\left[{1\over4}[-\ln\psi]^{\prime\prime2}+
{\chi_+(\theta)\chi_+^{\prime\prime}(\theta)
-\chi_+^{\prime2}(\theta)}-2
\chi_+^{2}(\theta)[-\ln\psi]^{\prime\prime}\right]\nonumber\\
&&-{1\over4}(-\psi^2(\theta)
[\ln\psi]^{\prime\prime})^{\alpha^\prime t}
[-\ln\psi]^{\prime\prime} +{1\over4\sin^2\theta}\bigg)\label{reggecorrection+}\\
\Sigma^{-}&=&-{8g^2\alpha^{\prime2-D/2}
\over(8\pi^2)^{D/2}}
\int_0^1{dq\over q}2^4\left({-\pi\over\ln q}\right)^{(10-D)/2}
(1+w^{1/2})^{10-D-S}{\prod_n(1+q^{2n})^{8}\over\prod_n(1-q^{2n})^{8}}\nonumber
\\
&&\hskip-.28in \int_0^{\pi} d\theta\bigg((-\psi^2(\theta)
[\ln\psi]^{\prime\prime})^{\alpha^\prime t}
{\alpha^\prime t\over[\ln\psi]^{\prime\prime}}
\left[{1\over4}[-\ln\psi]^{\prime\prime2}+
{\chi_-(\theta)\chi_-^{\prime\prime}(\theta)
-\chi_-^{\prime2}(\theta)}-2
\chi_-^{2}(\theta)[-\ln\psi]^{\prime\prime}\right]\nonumber\\
&&-{1\over4}(-\psi^2(\theta)
[\ln\psi]^{\prime\prime})^{\alpha^\prime t}
[-\ln\psi]^{\prime\prime} +{1\over4\sin^2\theta}\bigg)
\label{reggecorrection-}
\eea
We have made
the Neveu-Scherk subtraction in both $\Sigma^\pm$ in order that the
$\theta$ integrals are explicitly finite:
\bea
\psi\to\psi_C=\sin\theta,\qquad
\chi^+&\to&\chi_C={1\over2\sin\theta}\nonumber\\
-\psi_C^2[\ln\psi_C]^{\prime\prime}&=&1\nonumber\\
{1\over4}(1+\alpha^\prime t)[-\ln\psi_C]^{\prime\prime}
+\alpha^\prime t{\chi_C\chi_C^{\prime\prime}-\chi_C^{\prime\prime}
\over[-\ln\psi_C]^{\prime\prime}}-2\alpha^\prime t\chi_C^2
&=&{1\over4\sin^2\theta}
\eea
In fact these subtraction terms are actually zero with the GNS
regulator:
\bea
\int_0^{2\pi}d\theta (\sin\theta/2)^{P^2-2}={\Gamma(1/2)\Gamma((P^2-1)/2)
\over\Gamma(P^2/2)}
\to0,\qquad {\rm for}~P\to0.
\label{subtraction}
\eea
Furthermore as pointed out by Neveu and Scherk \cite{neveuscherkrenorm},
when $t=0$ the $\theta$ integral of the last terms in
the expressions for $\Sigma^\pm$ gives zero:
\bea
\int_0^\pi d\theta \left([\ln\psi]^{\prime\prime}+{1\over\sin^2\theta}\right)
&=&\left([\ln\psi]^{\prime}-\cot\theta\right)|_0^\pi
=\left(\sum_{n=1}^\infty {4q^{2n}\sin2\theta\over
1-2q^{2n}\cos2\theta+q^{4n}}\right)\bigg|_0^\pi=0
\eea
It will be convenient to  replace each $\sin^{-2}\theta$ term by 
$[-\ln\psi]^{\prime\prime}$. In addition we call the quantities in square
brackets in (\ref{reggecorrection+}) and (\ref{reggecorrection-}) 
$X^\pm$, so we can rewrite 
these two equations as
\bea
\Sigma^{+}&=&-{8g^2\alpha^{\prime2-D/2}
\over(8\pi^2)^{D/2}}
\int_0^1{dq\over q^2}\left({-\pi\over\ln q}\right)^{(10-D)/2}
(1-w^{1/2})^{10-D-S}{\prod_r(1+q^{2r})^{8}\over\prod_n(1-q^{2n})^{8}}
\nonumber\\
&&\hskip-.28in  \int_0^{\pi} d\theta\bigg((-\psi^2(\theta)
[\ln\psi]^{\prime\prime})^{\alpha^\prime t}
{\alpha^\prime t\over[\ln\psi]^{\prime\prime}}X^+
-{1\over4}[(-\psi^2(\theta)
[\ln\psi]^{\prime\prime})^{\alpha^\prime t}-1]
[-\ln\psi]^{\prime\prime}\bigg)\label{reggecorrectionfinal+}\\
\Sigma^{-}&=&-{8g^2\alpha^{\prime2-D/2}
\over(8\pi^2)^{D/2}}
\int_0^1{dq\over q}2^4\left({-\pi\over\ln q}\right)^{(10-D)/2}
(1+w^{1/2})^{10-D-S}{\prod_n(1+q^{2n})^{8}\over\prod_n(1-q^{2n})^{8}}
\nonumber\\
&&\hskip-.28in \int_0^{\pi} d\theta\bigg((-\psi^2(\theta)
[\ln\psi]^{\prime\prime})^{\alpha^\prime t}
{\alpha^\prime t\over[\ln\psi]^{\prime\prime}}X^- -{1\over4}[(-\psi^2(\theta)
[\ln\psi]^{\prime\prime})^{\alpha^\prime t}-1]
[-\ln\psi]^{\prime\prime}\bigg)
\label{reggecorrectionfinal-}
\eea
Written this way $\Sigma^\pm$ formally vanish as $t\to0$, which is
expected since the massless open string state is a gauge particle
and should remain massless.
The corrected Regge trajectory is
\bea
\alpha(t)&=&1+\alpha^\prime t +{1\over2}(\Sigma^+-\Sigma^-)\equiv
1+\alpha^\prime t + \Sigma(t)
\label{trajectory}
\eea
We close this section by noting a dramatic simplification of $X^\pm$:
\bea
X^+&=&{1\over4}\theta_4(0)^4\theta_3(0)^4-{{\bfs E}\over\pi}
\theta_4(0)^4\theta_3(0)^2+{{\bfs E}^2\over\pi^2}\theta_3(0)^4
=4q+O(q^2)\label{X+}\\
X^-&=&-{1\over4}\theta_4(0)^4\theta_3(0)^4+{{\bfs E}^2\over\pi^2}\theta_3(0)^4
=O(q^2)\label{X-}\\
{\bfs E}&=&{\pi\over6\theta_3(0)^2}\left({\theta_3(0)^4+\theta_4(0)^4}
-{\theta_1^{\prime\prime\prime}(0)\over\theta_1^\prime(0)}\right)
={\pi\over2}-2\pi q +O(q^2)
\eea
where we have shown on the extreme right of each equation its small
$q$ behavior.
Remarkably, $X^\pm$ turn out to be independent of $\theta$!
The most efficient way to show this is to express $\chi_\pm$ and
$-[\ln\psi]^{\prime\prime}$ in terms of the Jacobian elliptic
functions sn, cn, and dn. Then one exploits the many identities
these functions and their derivatives satisfy \cite{bateman}.
For later use we list the expansions
\bea
\theta_3(0)&=&\prod_n(1-q^{2n})\prod_r(1+q^{2r})^2\\
\theta_4(0)&=&\prod_n(1-q^{2n})\prod_r(1-q^{2r})^2\\
{\theta_1^{\prime\prime\prime}(0)\over\theta_1^\prime(0)}
&=&-1+24\sum_n {q^{2n}\over(1-q^{2n})^2}
\eea
where sums over $n$ are over positive integers and those over $r$
are over half odd integers.
\section{Small $t$ and the Field Theory Limit}\label{smallpsection}
The field theory limit of string theory is controlled by the zero
slope limit $\alpha^\prime\to0$. In open string theory, this
entails analyzing the behavior of physical quantities at
low momentum $\alpha^\prime \, k_l \! \cdot \! k_m \ll1$. In this section,
we wish to study the small $t$ behavior of the open string Regge
trajectory. The one loop correction to this trajectory 
should reflect the one-loop Reggeization of the gluon in gauge theory. 

The part of $\Sigma(t)$ analytic in $t$ (i.e. integer powers)
receives contributions from the
whole range of $q$, but non-analytic behavior in $\Sigma$ at $t=0$
is produced by integration of $q$ near 1. 
Thus it is convenient in this section to transform to the 
$w$ variables $\ln w=2\pi^2/\ln q
=-2\pi i/\tau$,
via the Jacobi imaginary transformation, and our focus will be
on the small $w$ contribution to $\Sigma$. 
The measure change is 
$q^{-1}dq\to 2\pi^2(w\ln^2w)^{-1}dw$, and the partition functions
change according to
\bea
P_+&\equiv&q^{-1}(1-w^{1/2})^{10-D-S}{\prod_r(1+q^{2r})^8\over\prod_n(1-q^{2n})^8}
=\left({2\pi\over\ln w}\right)^4w^{-1/2}(1-w^{1/2})^{10-D-S}
{\prod_r(1+w^{r})^8\over\prod_n(1-w^{n})^8}\label{partition+}\\
P_-&\equiv&2^4(1+w^{1/2})^{10-D-S}{\prod_n(1+q^{2n})^8\over\prod_n(1-q^{2n})^8}
=\left({2\pi\over\ln w}\right)^4w^{-1/2}(1+w^{1/2})^{10-D-S}
{\prod_r(1-w^{r})^8\over\prod_n(1-w^{n})^8}\label{partition-}
\eea
from which we find the small $w$ behavior
\bea
\left({-\pi\over\ln q}\right)^{5-D/2}{dq\over q}{P_++P_-\over2}
&\sim&{dw\over2w}\left({-2\pi\over\ln w}\right)^{1+D/2}
w^{-1/2}\\
\left({-\pi\over\ln q}\right)^{5-D/2}
{dq\over q}{P_+-P_-\over2}&\sim&(D+S-2){dw\over2w}
\left({-2\pi\over\ln w}\right)^{1+D/2}
\eea
Next we turn to the Jacobi transforms of the $\theta$ dependent factors.
\bea
\theta_1\left({i\theta\ln w\over 2\pi},\sqrt{w}\right)
&=&-i\left({-2\pi\over\ln w}\right)^{1/2}\exp\left\{{-\theta^2\ln w\over2\pi^2}
\right\}\theta_1(\theta,q)\label{jacobitheta1}\\
\theta_1^\prime(0,\sqrt{w})&=& 
\left({-2\pi\over\ln w}\right)^{3/2}\theta_1^\prime(0,q)
\label{jacobitheta1prime}\\
\psi(\theta,q)&=&{\theta_1(\theta,q)\over\theta^\prime(0)}
=i{-2\pi\over\ln w }\exp\left\{{\theta^2\ln w\over2\pi^2}
\right\}{\theta_1\left({i\theta\ln w/ 2\pi},\sqrt{w}\right)\over
\theta_1^\prime(0,\sqrt{w})}\nonumber\\
&=&{\pi\over-\ln w }\exp\left\{-{\theta(\pi-\theta)\ln w\over2\pi^2}
\right\}(1-w^{\theta/\pi})(1-w^{1-\theta/\pi})\nonumber\\
&&\hskip2in\prod_{n=1}^\infty{(1-w^{n+\theta/\pi})
(1-w^{n+1-\theta/\pi})\over(1-w^n)^2}\label{jacobipsi}\\
-{\partial^2\over\partial\theta^2}\ln\psi&=&{-\ln w\over\pi^2}
+{\ln^2 w\over\pi^2}\sum_{n=0}^\infty
\left[{w^{n+\theta/\pi}\over(1-w^{n+\theta/\pi})^2}
+{w^{n+1-\theta/\pi}\over(1-w^{n+1-\theta/\pi})^2}\right]
\label{jacobipsiprime}
\eea
We have written $\psi$ and its double logarithmic derivative in
a way that is manifestly symmetric under $\theta\to\pi-\theta$.
Continuing with our list of Jacobi transforms,
\bea
{\bfs E}&=&{-\ln w\over12\theta_3^2(0,\sqrt{w})}
\left(\theta_3^4(0,\sqrt{w})+\theta_2^4(0,\sqrt{w})
+{\theta_1^{\prime\prime\prime}(0.\sqrt{w})\over
\theta_1^\prime(0,\sqrt{w})}-{12\over\ln w}\right)\label{jacobie}\\
X^+&=&\left(-{\ln w\over2\pi}\right)^4{\theta_3^4(0,\sqrt w)
\theta_2^4(0,\sqrt w)\over4}-{{\bfs E}\over\pi}
\left(-{\ln w\over2\pi}\right)^3{\theta_3^2(0,\sqrt w)
\theta_2^4(0,\sqrt w)\over4}\nonumber\\
&&+{{\bfs E}^2\over\pi^2}\left(-{\ln w\over2\pi}\right)^2\theta_3^4(0,\sqrt{w})
\label{jacobix+}\\
X^-&=&-\left(-{\ln w\over2\pi}\right)^4{\theta_3^4(0,\sqrt w)
\theta_2^4(0,\sqrt w)\over4}+{{\bfs E}^2\over\pi^2}
\left(-{\ln w\over2\pi}\right)^2\theta_3^4(0,\sqrt{w})\\
X^+-X^-&=&2\left(-{\ln w\over2\pi}\right)^4{\theta_3^4(0,\sqrt w)
\theta_2^4(0,\sqrt w)\over4}-{{\bfs E}\over\pi}
\left(-{\ln w\over2\pi}\right)^3{\theta_3^2(0,\sqrt w)
\theta_2^4(0,\sqrt w)\over4}\label{jacobix-}\\
X^++X^-&=&-{{\bfs E}\over\pi}
\left(-{\ln w\over2\pi}\right)^3{\theta_3^2(0,\sqrt w)
\theta_2^4(0,\sqrt w)\over4}+2{{\bfs E}^2\over\pi^2}
\left(-{\ln w\over2\pi}\right)^2\theta_3^4(0,\sqrt{w})
\eea
Finally we need the small $w$ behavior of these combinations of 
$\theta$ functions:
\bea
\theta_3(0,\sqrt{w})&=&\prod_n(1-w^n)\prod_r(1+w^r)^2\sim 1+2w^{1/2}+O(w)
\nonumber\\
\theta_2(0,\sqrt{w})&=&2w^{1/8}\prod_n(1-w^n)\prod_r(1+w^r)^2
\sim 2w^{1/8}(1+O(w))\nonumber\\
{\theta_1^{\prime\prime\prime}(0,\sqrt{w})\over\theta_1^{\prime}(0,\sqrt{w})}
&=&-1+24\sum_n {w^n\over(1-w^n)^2}\sim -1+O(w),\qquad
{\bfs E}\sim 1+O(w^{1/2}\ln w)\\ 
X^+-X^-&\sim& 8w^{1/2}\left({\ln w\over2\pi}\right)^4\left(1+O\left({1\over\ln w}\right)\right),\qquad X^++X^-\sim{2\over\pi^2}\left({\ln w\over2\pi}\right)^2(1+O(w^{1/2}))
\eea
$X^+$ and $X^-$ enter the integrand of $\Sigma_+-\Sigma_-$ in the
combination
\bea
 P_+ X^+ -P_-X^-&=&(X^+-X^-){P_++P_-\over2}+(X^++X^-){P_+-P_-\over2}\\
&\sim&8w^{1/2}\left({\ln w\over2\pi}\right)^4{P_++P_-\over2}
+{2\over\pi^2}\left({\ln w\over2\pi}\right)^2{P_+-P_-\over2}\\
\left({-\pi\over\ln q}\right)^{5-D/2}{dq\over q}( P_+ X^+ -P_-X^-)
&\sim&{dw\over w}\left({-2\pi\over\ln w}\right)^{-3+D/2}
\left[4
+{D+S-2\over\pi^2}\left(2\pi\over{\ln w}\right)^{2}\right]
\eea 
Thus the contribution of the small $w$ region to the correction to the
Regge trajectory, which can produce nonanalytic behavior as $t\to0$, is
\bea
\Sigma&\approx&-{8g^2\alpha^{\prime2-D/2}\over(8\pi^2)^{D/2}}\int_0^\delta
{dw\over w}\left({-2\pi\over\ln w}\right)^{-3+D/2}
\int_0^\pi d\theta\bigg((-\psi^2(\theta)
[\ln\psi]^{\prime\prime})^{\alpha^\prime t}
{\alpha^\prime t\over[\ln\psi]^{\prime\prime}}\left[2+{D+S-2\over2\pi^2}\left(2\pi\over{\ln w}\right)^{2}\right]\nonumber\\
&&-{D+S-2\over8}\left(2\pi\over{\ln w}\right)^4[(-\psi^2(\theta)
[\ln\psi]^{\prime\prime})^{\alpha^\prime t}-1]
[-\ln\psi]^{\prime\prime}\bigg)\; .
\label{sigmadelta}
\eea
To discuss the small $t$ behavior of $\Sigma(t)$, we examine the $t$
dependent factor of the small $w$ integrand. We set $w=e^{-T}$ and
$\theta=\pi x$, and
note that at large $T>e^{1/\delta}$
\bea
-\psi^2(\theta)
[\ln\psi]^{\prime\prime}&\sim&e^{x(1-x)T}(1-e^{-xT})^2(1-e^{-(1-x)T})^2
\left[{1\over T}
+{e^{-xT}\over(1-e^{-xT})^2}
+{e^{-(1-x)T}\over(1-e^{-(1-x)T})^2}\right]
\label{psilogpsi}
\eea
This quantity is raised to the power $\alpha^\prime t=-\alpha^\prime |t|$
since we limit ourselves to $t<0$, where the integral representation
of $\Sigma$ is valid. The factor 
$e^{-x(1-x)T\alpha^\prime|t|}$ limits the quantity
$x(1-x)T<1/\alpha^\prime|t|$. Nonanalytic behavior in
$t$ as $t\to0$ can be produced by integration over the region
$1\ll\Lambda<x(1-x)T<\infty$. In this region $w^{\theta/\pi}
=e^{-xT}\ll1$ and $w^{1-\theta/\pi}
=e^{-(1-x)T}\ll1$, so the $t$ dependent factor in the integrand
reduces to
\bea
(-\psi^2(\theta)[\ln\psi]^{\prime\prime})^{-\alpha^\prime|t|}
&\sim& T^{\alpha^\prime|t|}e^{-x(1-x)T\alpha^\prime|t|}
\eea
Then the small $w$ (or large $T$) contribution to $\Sigma$ simplifies to
\bea
\Sigma&\approx&-{8g^2\alpha^{\prime2-D/2}\over(8\pi^2)^{D/2}}
\int_{\Lambda/x(1-x)}^\infty
{dT}\left({2\pi\over T}\right)^{-2+D/2}
\int_0^1\pi dx \bigg(
{-\pi\alpha^\prime t}T^{\alpha^\prime|t|}
e^{-x(1-x)T\alpha^\prime|t|}\nonumber\\
&&- {D+S-2\over4\pi}\left({2\pi\over T}\right)^2
\left[T^{\alpha^\prime|t|}
e^{-x(1-x)T\alpha^\prime|t|}-1\right]\bigg)\nonumber\\
&\approx&-{8g^2\alpha^{\prime2-D/2}\over(4\pi)^{D/2}}
\bigg(
{-\alpha^\prime t\over4}
{\Gamma(-2+\alpha^\prime t+D/2)^2\over
\Gamma(D-4+2\alpha^\prime t)}\int_{\Lambda}^\infty
{du}{u}^{2-\alpha^\prime t-D/2}e^{-u\alpha^\prime|t|}
\nonumber\\
 && 
-{D+S-2\over4}
\int_{\Lambda}^\infty
{du}{u}^{-D/2}\left[{\Gamma(\alpha^\prime t+D/2)^2\over
\Gamma(D+2\alpha^\prime t)}u^{-\alpha^\prime t}
e^{-u\alpha^\prime|t|}-{\Gamma(D/2)^2\over
\Gamma(D)}\right]\bigg)
\eea
The first integral is of the form 
(putting $\xi=\alpha^\prime|t|=-\alpha^\prime t$)
\bea
\int_\Lambda^\infty du u^ae^{-u\xi}&=&\xi^{-a-1}\int_{\xi\Lambda}^\infty
du u^ae^{-u}\sim \xi^{-a-1}\Gamma(a+1),\qquad a+1>0
\eea
If $a+1\leq0$, a few integration by parts shows that
\bea
\int_\Lambda^\infty du u^ae^{-u\xi}&=&\xi^{-a-1}\int_{\xi\Lambda}^\infty
du u^ae^{-u}\sim \xi^{-a-1}\Gamma(a+1)+{\cal P}(\xi,\Lambda)e^{-\xi\Lambda}
\label{asymptint}\eea
where ${\cal P}$ is a polynomial in $\xi$ (and hence a polynomial
in $t$) with $\Lambda$ dependent 
coefficients. In the expression for $\Sigma$ 
the term involving ${\cal P}$ is multiplied by a function analytic
at $t=0$ and so contributes integer powers of
$t$ to $\Sigma$. Thus the nonanalytic fractional power,
though always present, dominates
the behavior only if $a+1>0$. For the first integral this means $D<6$.

The subtraction term in the second integral is finite provided
$D>2$, which is not a serious restriction. The analysis of that
integral again leads to an expression of the form (\ref{asymptint}),
with the subtraction term simply cancelling the
constant term in the power series arising from ${\cal P}(\xi,\Lambda)$. 
Collecting the results for both integrals
we find the small $t$ behavior
\bea
\Sigma&\sim&-{2g^2\alpha^{\prime2-D/2}\over(4\pi)^{D/2}}\bigg(
{\Gamma(-2+\alpha^\prime t+D/2)^2\over
\Gamma(D-4+2\alpha^\prime t)}(-\alpha^\prime t)^{\alpha^\prime t-2+D/2}
\Gamma(3-\alpha^\prime t-D/2)
\nonumber\\
 && 
-(D+S-2)
{\Gamma(\alpha^\prime t+D/2)^2\over
\Gamma(D+2\alpha^\prime t)}(-\alpha^\prime t)^{\alpha^\prime t-1+D/2}
\Gamma(1-\alpha^\prime t-D/2)\bigg)+O(\alpha^\prime t)\\
&\sim&-{2g^2\alpha^{\prime2-D/2}\over(4\pi)^{D/2}}
{\Gamma(-2+D/2)^2\over
\Gamma(D-4)}(-\alpha^\prime t)^{-2+D/2}
\Gamma(3-D/2)+O(\alpha^\prime t \ln(-\alpha^\prime t))\; .
\label{sigmasmallw}
\eea
In the first line, we show the second term proportional to $(D+S-2)$, which 
is down by a factor of $\alpha^\prime t$ compared to the first term to
stress that the number of massless scalars $S$ does not figure in the
leading small $t$ behavior. With the $O(\alpha^\prime t \ln(-\alpha^\prime t))$
term we stress that the displayed first term is dominant only for
$D<6$, and when $D>6$ it is only significant when $D/2$ is not
an integer. The contributions to $\Sigma$ from $w$ away from zero by a
finite amount are analytic in $t$, i.e. a series of integer powers of $t$ 
starting with the first power, since $\Sigma(0)=0$.  
To properly regulate infrared divergences
(which would, among other things, 
invalidate the integral representation for $\Sigma$),
we would also need to stipulate that $D>4$, so in practice this formula gives
the dominant small $t$ behavior only in the range $4<D<6$.

The leading small $t$  
behavior $(-\alpha^\prime t)^{(D-4)/2}$ we have demonstrated 
is precisely that of a 
$D$-dimensional gauge theory: the one loop correction of which 
is consistent with
Regge behavior with a trajectory $\alpha(t)=1+C(-t/\mu^2)^{(D-4)/2}$.
As $D\to4$ from above the pole at $D=4$ shows that a $\ln(-\alpha^\prime t)$
dependence remains:
\bea
\Sigma_{D\to4}\sim -{g^2\over4\pi^2}\left[{2\over D-4}+\ln(-\alpha^\prime t)
\right]
\eea
Consulting known results for one loop gauge theory diagrams
\cite{kunsztst}, as explained and developed in 
\cite{thornresir,chakrabartiqt,chakrabartiqt2},
we find that the small $t$ behavior of the open string trajectory
agrees exactly with that inferred from conventional one loop
gauge theory calculations.
%
%
\section{Large $t$ Behavior of $\Sigma(t)$}\label{largepsection}
The large $t$ behavior of the trajectory function is dominated by 
the small $q$ region. Thus it is more convenient to write the 
integral in this variable. Combining equations 
(\ref{reggecorrectionfinal+}) and (\ref{reggecorrectionfinal-}) we have
\bea \label{sigmalargep}
\Sigma(t) \eqe -\frac{4g^2\alpha'^{2-D/2}}{(8\pi^2)^{D/2}} \int_0^1 
\frac{dq}{q} \left(\frac{-\pi}{\ln q}\right)^{(10-D)/2} 
\int_0^{\pi} d\theta 
\left((-\psi^2(\theta)
[\ln \psi]^{\prime\prime})^{\alpha' t}
\frac{\alpha' t}{[\ln \psi]^{\prime\prime}}(P_+X^+-P_-X^-)\right.\nonumber\\
&& \left.-\frac{1}{4}(P_+-P_-)\left[(-\psi^2(\theta)
[\ln \psi]^{\prime\prime})^{\alpha't}-1\right] 
[-\ln \psi]^{\prime\prime} \right)
\eea
For later convenience, we write the partition functions as
\bea
\label{partw}
P_+ &=& 16 (1-w^{1/2})^{10-D-S} \frac{\theta_3(0)^4}{\theta_1'(0)^4}
\nonumber \\
P_- &=& 16 (1+w^{1/2})^{10-D-S} \frac{\theta_2(0)^4}{\theta_1'(0)^4}
\eea
where we have used the fact that
\bea
\frac{\prod_{r}^{\infty}(1+q^{2r})^8}{\prod_{n}^{\infty}(1-q^{2n})^8} 
= 16 q \frac{\theta_3(0)^4}{\theta_1'(0)^4} \quad , \quad 
\frac{\prod_{n}^{\infty}(1+q^{2n})^8}{\prod_{n}^{\infty}(1-q^{2n})^8} 
= 16 q \frac{\theta_2(0)^4}{\theta_1'(0)^4}
\eea
Expanding about $q=0$ we have:
\bea
P_+-P_- \simeq P_+ \sme q^{-1} \left(\frac{-\pi^2}{\ln q} \right)^{10-D-S}\\
P_+X^+-P_-X^- \sme 4 \left(\frac{-\pi^2}{\ln q} \right)^{10-D-S}\\
\left[-\ln \psi\right]'' &=& \csc^2 \theta + \mathcal{O}(q^2)\\
\log (-\psi^2 (\log \psi)^{\prime\prime}) &=& 
16 q^2 \sin^4 \theta + \mathcal{O}(q^4)
\eea
The small $q$ contribution to the trajectory function 
(\ref{sigmalargep}) becomes
\bea\
\label{sigmalargep2}
\Sigma(t) \sme \frac{16g^2\alpha'^{2-D/2}}{(8\pi^2)^{D/2}} 
\alpha' t \int_0^{\delta} \frac{dq}{q} 
\left(\frac{-\pi}{\ln q}\right)^{25-5D/2-2S} 
\int_0^{\pi} d\theta \, e^{-\alpha' |t| 16 q^2 \sin^4 \theta} 
\sin^2 \theta\nonumber\\
&& + \, \frac{g^2\alpha'^{2-D/2}}{(8\pi^2)^{D/2}}  
\int_0^{\delta} \frac{dq}{q^2} \left(\frac{-\pi}{\ln q}\right)^{25-5D/2-2S} 
\int_0^{\pi} d\theta \left[e^{-\alpha' |t| 16 q^2 \sin^4 \theta} -1 \right] 
\csc^2 \theta
\eea
Although the presence of the explicit factor of $t$ in front of 
the first integral above makes this term the dominant one, 
a careful analysis shows that this term is larger than the second 
one by a factor of $\sqrt{t}$, not by a factor of $t$. Therefore 
we study the expression
\bea
I \equiv \int_0^{\delta} \frac{dq}{q} \left(\frac{-1}{\ln q}\right)^a 
\int_0^{\pi} d\theta \, \sin^2 \theta \, e^{-16 q^2 |\alpha' t| \sin^4 \theta}
\eea
in the limit $|\alpha' t| \to \infty$ where $a \equiv 25-5D/2-2S$. 
Performing the change of variable $u=-\ln q$ we have
\bea
I \equiv \int_{-\ln \delta}^{\infty} du \, \left(\frac{1}{u}\right)^a 
\int_0^{\pi} d\theta \, \sin^2 \theta \, 
\exp{\left[{-\exp{\left[\ln(16 |\alpha' t| \sin^4 \theta)-2u\right]}}\right]}
\eea
followed by the change $u=\xi \ln (16 \alpha' |t| \sin^4 \theta)$, yields
\bea
I \equiv  \int_0^{\pi} d\theta \, \sin^2 \theta f(\theta)^{1-a} \, 
\int_{{\ln \delta}/{f(\theta)}}^{\infty} d\xi \,\, \xi^{-a} 
\exp{\left[{-\exp{\left[(1-2\xi)f(\theta)\right]}}\right]}
\eea
where $f(\theta) \equiv \ln (16 \alpha' |t| \sin^4 \theta)$. 
At fixed $\theta$, the exponential factor restricts the integration 
over $\xi$ to the range $[1/2,\infty)$ as $|t| \to \infty$, and since there 
are no singularities coming from the end points of the 
$\theta$ integration we have
\bea
I \sme  \left(\frac{1}{\ln |\alpha' t|}\right)^{a-1} 
\int_{0}^{\pi} 
d\theta \, \sin^2 \theta \, \int_{1/2}^{\infty} d\xi \,\, \xi^{-a} \\
\sme \frac{\pi}{2} \left(\frac{1}{\ln |\alpha' t|}\right)^{a-1} 
\frac{2^{a-1}}{a-1}
\eea
Putting this result into (\ref{sigmalargep2}) we finally have
\bea\label{largetestimate}
\Sigma(t) \sme g^2 \alpha' \frac{(8\pi^2\alpha')^{1-D/2}}{24-5D/2-2S} 
\alpha' t \left(\frac{2\pi}{\ln |\alpha' t|}\right)^{24-5D/2-2S}
\eea
as $|t| \to \infty$.
%
%
\section{Numerical Calculations and Graphics}\label{numerics}
We present here a numerical analysis of the trajectory function 
$\Sigma (t)$ in both limits studied, large $t$ and small $t$ as 
well as in the full range between these two asymptotic regions. 
We will see that the numerical computation of the integral representation 
of the trajectory function  $\Sigma (t)$ matches well with the asymptotic 
analytical predictions described in sections \ref{smallpsection} and 
\ref{largepsection}.

\begin{figure}[ht!]
\centering
\includegraphics[scale=0.70]{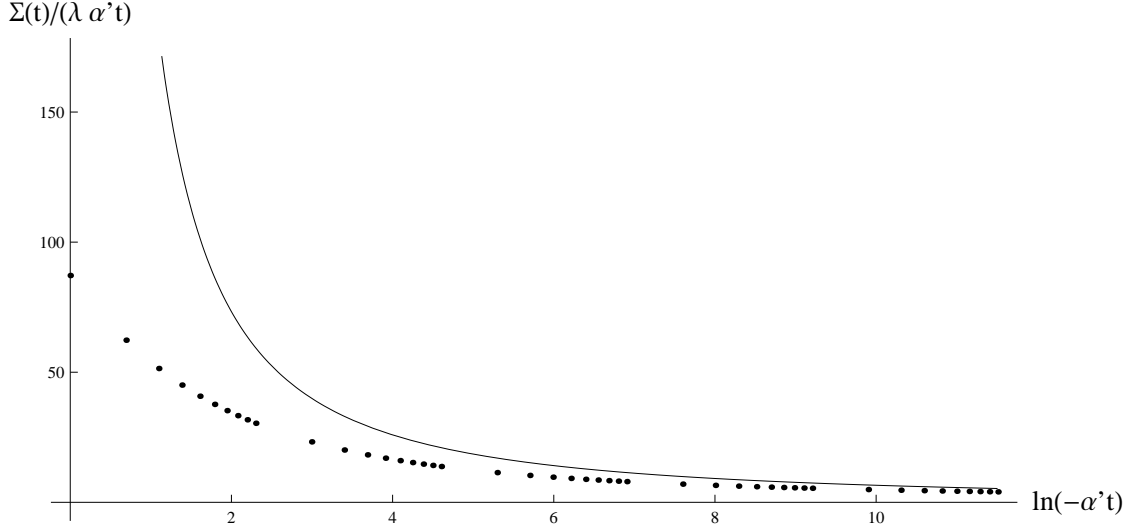}
\caption{The dots correspond to the direct numerical 
integration of $\Sigma(t)/(\lambda t)$ while the solid line is 
the predicted behavior at large $|t|$ both as a function of 
$\ln(-\alpha't)$. At large values of $\alpha' |t|$ the 
numerical integration approaches the predicted behavior from below.}
\label{fig:largetlog}
\end{figure}

\begin{figure}[ht!]
\centering
\includegraphics[scale=0.70]{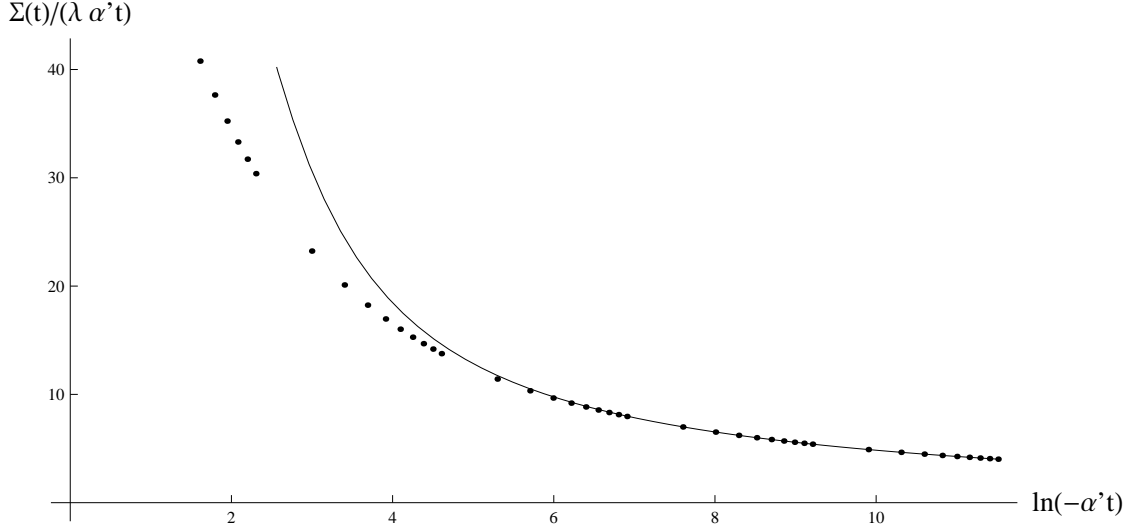}
\caption{The fit of the leading and three subleading 
corrections to the data points is presented as the solid line. 
Here we see that the agreement between the data points and the 
fit increases for larger values of $\alpha't$ as expected, and 
it is already good starting at $\ln(\alpha't) \sim 6$ since we 
have included subleading corrections.}
\label{fig:fitted}
\end{figure}

\begin{figure}[ht!]
\centering
\includegraphics[scale=0.65]{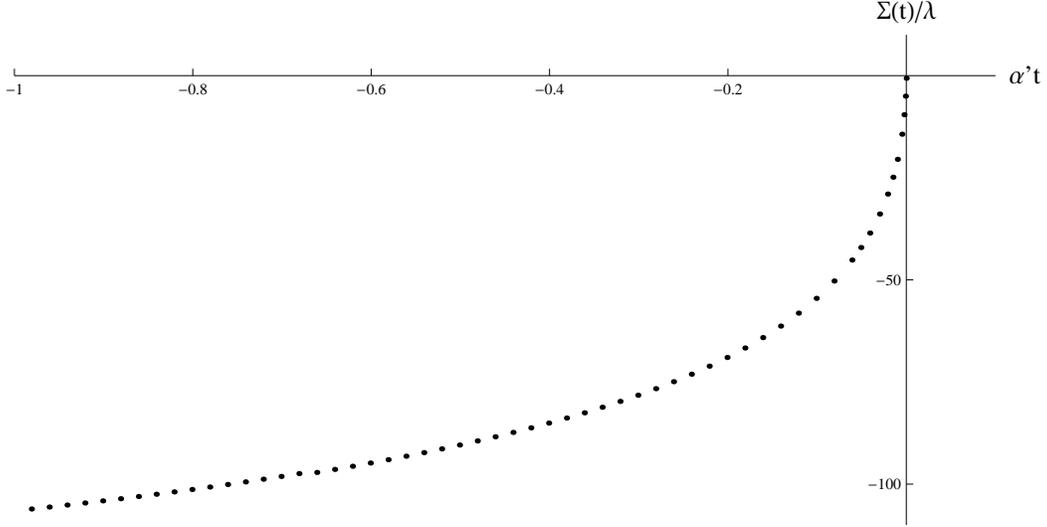}
\caption{The small $t$ behavior is shown for $D=5$. 
We expect $\Sigma(t)$ to go to zero with infinite slope 
which can be appreciated from this figure.}
\label{fig:smallt1}
\end{figure}

At large $t$ we expect $\Sigma(t)$ to behave like (\ref{largetestimate}). 
As a numerical example, let us consider the case 
when there is the maximum number of scalars circulating in the loop $S=10-D$, 
in $D=5$ i.e. open strings ending on a D4-brane. We expect at large $t$:
\bea
\Sigma(t)_{lead} \sme \lambda \alpha' t \frac{4\pi^2}{3} 
\left(\frac{2\pi}{\ln |\alpha't|}\right)^{3/2} 
\eea
where $\lambda \equiv {4g^2 \alpha'^{2-D/2}}/{(8\pi^2)^{D/2}}$. For 
plotting convenience, we show in Figure \ref{fig:largetlog} both the 
numerical evaluation of  ${\Sigma(t)}/{\lambda \alpha' t}$ directly 
from equation (\ref{sigmalargep}) represented by the dots, and we show the leading estimate
\bea
\label{leading}
\frac{\Sigma(t)_{lead}}{\lambda \, \alpha' t}  \sim 
\frac{4\pi^2}{3}\left(\frac{2\pi}{\ln |\alpha' t|}\right)^{3/2}
\eea
by the solid line.
From the figure 
we see that the numerical evaluation of the full trajectory 
function (\ref{sigmalargep}) approaches the predicted behavior at large $|t|$,
although the approach is very slow due to the logarithmic dependence.
Due to computational limitations of the 
integration routine, the maximum value of $-\alpha' t$ for which 
the trajectory function could be numerically evaluated was of 
the order of $-\alpha' t \sim 10^5$ which corresponds to the upper 
limit $\ln(10^5)\sim 11.5$  in the horizontal axis shown in the figures.
Performing a fit of the exact numerical evaluation using 
the leading and the next three subleading corrections we obtain
\bea
\label{eqfit}
\frac{\Sigma}{\lambda \, \alpha' t} \sim 12.89 \left(\frac{2\pi}{\ln |\alpha' t|}\right)^{3/2} 
- 7.74 \left(\frac{2\pi}{\ln |\alpha' t|}\right)^{5/2} 
+ 4.84 \left(\frac{2\pi}{\ln |\alpha' t|}\right)^{7/2} - 0.85 \left(\frac{2\pi}{\ln |\alpha' t|}\right)^{9/2}
\eea
whose leading term is to be compared with the predicted leading behavior
\bea
\frac{\Sigma(t)_{lead}}{\lambda \, \alpha' t}  
\sim \frac{4\pi^2}{3}\left(\frac{2\pi}{\ln |\alpha' t|}\right)^{3/2} 
\sim 13.16 \left(\frac{2\pi}{\ln |\alpha' t|}\right)^{3/2}
\eea
which is in good agreement with the analytic prediction. 
Figure \ref{fig:fitted} 
shows the fitting function (\ref{eqfit}) together with the data points.\\
Now we turn to the small $t$ behavior. As has been noted 
in section 3, this is best studied in the $w$ variables 
since it is the $w \sim 0$ region that dominates the
nonanalytic behavior at small $t$. 
We need the exact form of $\Sigma(t)$ which, using the integral representation 
as a function of $w$ is:
\bea \label{sigmaw}
\Sigma(t) \eqe \lambda \int_0^1 \frac{dw}{2w} 
\left(\frac{-2\pi}{\ln w}\right)^{-3+D/2} 
\int_0^{\pi} d\theta \left((-\psi^2(\theta)
[\ln \psi]^{\prime\prime})^{\alpha' t}
\frac{\alpha' t}{[\ln \psi]^{\prime\prime}}(P_+X^+-P_-X^-)\right.\nonumber\\
&& \left.-\frac{1}{4}(P_+-P_-)\left[(-\psi^2(\theta)
[\ln \psi]^{\prime\prime})^{\alpha't}-1\right] 
[-\ln \psi]^{\prime\prime} \right)
\eea
To evaluate it numerically we need the exact 
$w$-dependent forms of the expressions 
listed in section \ref{smallpsection}, the partition functions in 
(\ref{partw}), and also:
\bea
[-\ln \psi]'' = \left(\frac{\ln w}{2\pi}\right)^2 \theta_3^2(0,\sqrt{w}) 
\theta_2^2(0,\sqrt{w}) 
\frac{ \theta_4^2(i \theta \ln w/2\pi,\sqrt{w})}
{\theta_1^2(i \theta \ln w/2\pi,\sqrt{w})} 
- \frac{2 {\bf E}}{\pi} \left(\frac{-\ln w}{2\pi}\right)\theta_3^2(0,\sqrt{w})
\eea
The numerical integration routine produces the plot in Figure 
\ref{fig:smallt1} for the range $\alpha't \in [-1,-0.01]$  
which suggests that $\Sigma(t)$ approaches zero with infinite slope 
as it should reading off from (\ref{sigmasmallw}). Nonetheless, 
we would like to see whether the numerics is really producing this 
behavior at small $t$. 
In order to do this, we need to zoom in near zero in Figure \ref{fig:smallt1} 
in which case we do the following: we separate out the small $w$ region as 
\bea
\int_0^1 dq [\cdots] = \int_0^{q_0} dq [\cdots] + \int_0^{w_0} dw [\cdots]
\eea
\begin{figure}[ht!]
\centering
\includegraphics[scale=0.65]{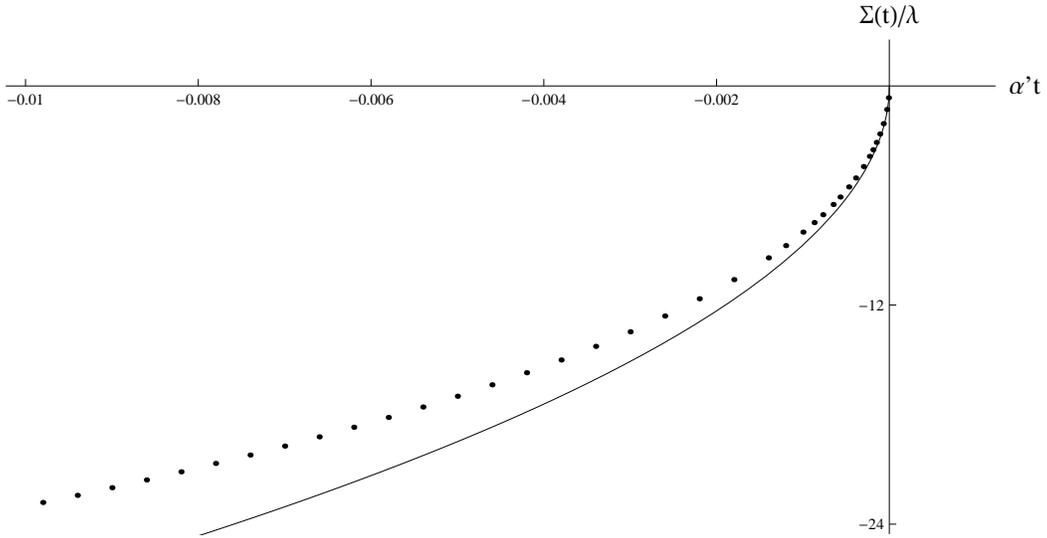}
\caption{Zoom in small $t$. The expected $\Sigma \sim (-\alpha' t)^{1/2}$ 
behavior near $t=0$ can be 
appreciated more clearly in this figure. The solid line is the 
predicted asymptotic behavior  $\Sigma(t)/\lambda 
\sim - 2^{3/2} \pi^4 (-\alpha' t)^{1/2}$ as $t \to 0 $ 
which matches well with the data points in this limit.}
\label{fig:zoomsmallt}
\end{figure}
If for instance, we take $q_0=0.8$, by means of $w=e^{2\pi^2/\ln q}$ we see
that the upper limit in the $w$ integration is $w_0 \sim 4 \times 10^{-39}$ 
which is very small and allows one to use the asymptotic expression 
(\ref{sigmadelta}) with $\psi$ and  $[-\ln \psi]^{\prime\prime}$
being also approximated by:
\bea
\psi \sme \frac{\pi}{T} e^{x/2(1-x)T}(1-e^{-xT})(1-e^{-(1-x)T})\\
\left[-\ln \psi\right]^{\prime\prime} \sme \frac{1}{\pi^2}  
\left[T+T^2 \left(\frac{e^{-xT}}{(1-e^{-xT})^2}+\frac{e^{-(1-x)T}}{(1-e^{-(1-x)T})^2}\right)\right]
\eea
as shown in equation (\ref{psilogpsi}). The numerical evaluation of 
(\ref{sigmadelta}) in this case is shown in Figure \ref{fig:zoomsmallt}. A fit of the data points including the
leading and subleading powers turns out to be
\bea
\frac{\Sigma}{\lambda} \sim -262.82 (-\alpha't)^{1/2} - 323.26 \, \alpha' t
\label{smalltcorrections}
\eea
\begin{figure}[ht]
\centering
\includegraphics[scale=0.70]{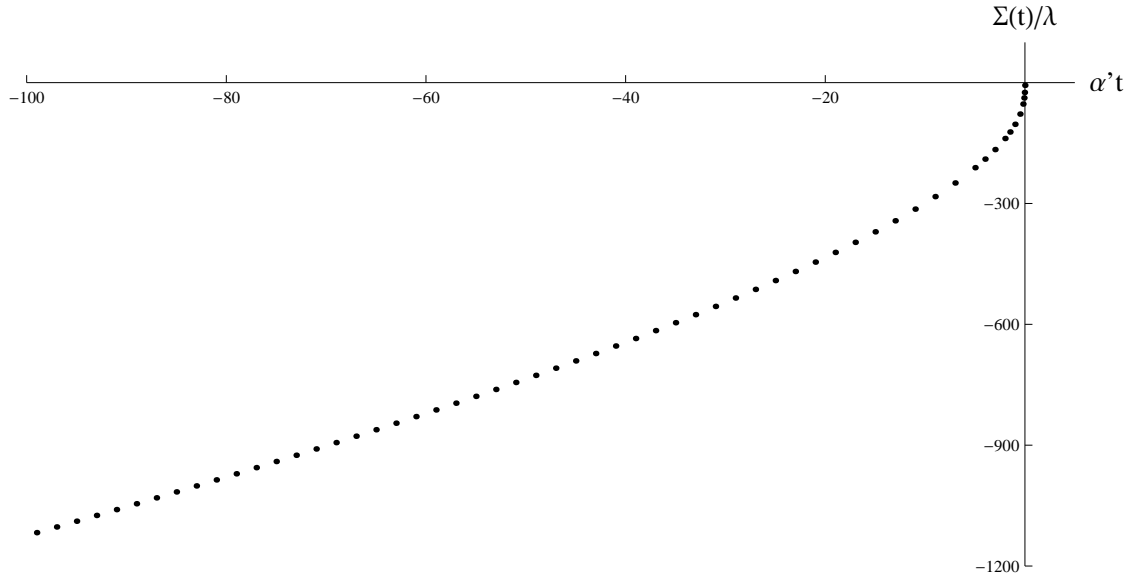}
\caption{A larger range that includes both 
large and small $t$ behavior is shown. In this plot it is 
possible to see the two asymptotic regions with some accuracy. 
The large $t$ region grows as $\sim t/(\ln t)^{3/2}$ as 
described in Section 5. Although it is not completely evident 
from this figure, $\Sigma(t)$ is going to zero with infinite 
slope as $(-\alpha't)^{1/2}$ in the region near $t=0$ 
(see Figure \ref{fig:zoomsmallt})}
\label{fig:fullrange1}
\end{figure} 
The analytic expression for the leading behavior is given by 
(\ref{sigmasmallw}) which in this case becomes
\bea
\frac{\Sigma}{\lambda} \sim - \frac{(2\pi)^{D/2}}{2} 
\frac{\Gamma(-2+D/2)^2}{\Gamma(D-4)} (-\alpha't)^{D/2-2} 
\Gamma(3-D/2) \sim - 2^{3/2} \pi^4 (-\alpha't)^{1/2}
\eea
for $D=5$. From here we see that we have good agreement 
with (\ref{smalltcorrections}) since $2^{3/2} \pi^4 = 275.52$.
We finish this section by showing a larger range in $t$ in which the 
two asymptotic regions $t \sim 0$ and $t\to -\infty$ can be visualized. 
Although it is not evident from the plot, Figure \ref{fig:fullrange1} shows the
two asymptotic regions we have described above i.e., 
$\Sigma \sim (-\alpha't)^{1/2}$ 
as $t \to 0$ and  $\Sigma \sim \alpha't / (\ln|\alpha't|)^{3/2}$ 
as $t \to -\infty$.
\section{Discussion and Conclusions}
\indent In this paper we have understood one more aspect about 
the relation between the dynamics of open strings and
their corresponding gauge field theory limit, namely 
we have extracted the leading Regge trajectory 
$\alpha(t)=1+\alpha't + \Sigma(t)$ of open strings ending on a 
stack of $N$ Dp-branes including the planar one-loop corrections.
When studying the $t \to 0$ limit, besides confirming that the 
correction goes to zero as a reflection of the masslessness of the open 
string gluon, we have also shown that it behaves as 
$\Sigma \sim -C g^2(\alpha' t)^{(D-4)/2}/(D-4)$ which is 
precisely the result obtained in $D$ dimensional gauge theory. 
In the $D\to 4$ case, which corresponds to a D3-brane where the 
dynamics of the open strings describe the 4-dimensional $SU(N)$ 
gauge-theory in 't Hooft's planar limit, the $t \to 0$ limit of $\Sigma(t)$ is
\bea
\Sigma_{D\to4}\sim -{g^2\over4\pi^2}\left[{2\over D-4}+\ln(-\alpha^\prime t)
\right]
\eea
which precisely matches the results known from one-loop 
gauge theory calculations. We have also studied the limit 
$t \to -\infty$ of $\Sigma(t)$ where we obtained:
\bea
\Sigma(t) \sme g^2 \alpha' \frac{(8\pi^2\alpha')^{1-D/2}}{24-5D/2-2S} 
\alpha' t \left(\frac{2\pi}{\ln |\alpha' t|}\right)^{24-5D/2-2S}
\eea
Since the maximum number of scalars circulating in the loop is 
$S_{\rm {max}}=10-D$  the 1-loop correction stays smaller than the 
tree trajectory in the entire range $-\infty<t<0$ as long as $D<8$.\\
\indent We have seen once again the usefulness of the 
GNS regulator\footnote{See Appendix A in \cite{thornsubcritical} 
for an application of this regulator in conventional field theory.} 
when dealing 
with the ``spurious'' divergences encountered in the 
integration over $\theta$. 
The subtraction (\ref{subtraction}) is very simple, and it is actually zero 
after analytic continuation to $P=0$.

\indent When we organized the integrand in the form of 
equations (\ref{reggecorrectionfinal+}) and (\ref{reggecorrectionfinal-}), 
the recognition that the quantities $X^{\pm}$ defined in (\ref{X+}) 
and (\ref{X-}) were independent of the integration variable $\theta$ 
significantly improved the numerical calculations performed in Section 7.\\
\indent There is still much work to be done. We are pursuing the idea that 
some aspects of QCD both perturbative and nonperturbative (such as confinement 
as an example of the latter) could be more tractable in the open string theory 
that has (large $N$) QCD as its low energy limit rather 
than in the field theory itself. 
One of the simplest open string theories that satisfies this requirement, and 
the one that we used in this article, is the even G-Parity sector of the 
Neveu-Schwarz model in 10 dimensions with the odd G-Parity 
states projected out to 
have a tachyon free spectrum. The end 
points of the open strings are required to be attached to a stack of $N$ 
Dp-branes as is customary. Since QCD does not contain massless scalars, 
we also  prevent them from circulating in the loop by projecting 
them out using the 
proposal in \cite{thornnonabelian}. However, our calculations also apply for 
Yang-Mills theories with massless scalars by simply 
choosing different values of 
the number of massless scalars $S$ in equations (\ref{reggecorrectionfinal+}) 
and (\ref{reggecorrectionfinal-}) as long as $0\leq S \leq 10-D$ where $D$ is 
the space-time dimensionality of the Dp-brane.
 \\
\indent Summarizing, we expect to learn much more from 
this model in the near future.
One immediate and straightforward follow-up to this work would be to 
repeat the analysis in the cases where the open strings 
end on D7 and D8 branes to encompass these two extra cases as well. 
In these cases there are extra subleading divergences that need to be taken 
care of (due to the emission of massless closed 
string states into the vacuum) which 
however can be absorbed into a renormalization of the 
Regge slope parameter $\alpha'$. 
Further studies of this model, such as an explicit expression free 
of spurious divergences for the complete 1-loop planar 
$M$-gluon amplitude, and the 
comparison of the low energy limits of these string amplitudes to the limiting 
field theory calculations  are tasks for the immediate future. We regard this 
work to be a further step in setting up the calculation for the 
complete sum of planar
open string multiloop diagrams, which we hope will shed light on 
nonperturbative issues of gauge theory.
\vskip14pt
\noindent\underline{ Acknowledgments}: 
We would like to thank Andr\'e Neveu for helpful discussions.
F.R. thanks Christoph Sachse for helpful discussions and Chris Pankow 
and Myeonghun Park for assistance with plotting.
This research was supported in part by the Department of 
Energy under Grant No. DE-FG02-97ER-41029.

\end{document}